\documentclass[aps,letterpaper,amssymb,twocolumn,pra]{revtex4}
\usepackage{bm}
\usepackage{amsmath}
\usepackage{amssymb}
\usepackage{times}
\usepackage{latexsym}
\usepackage{graphicx}
\usepackage{color}
\usepackage{subfigure}
\begin{document}
\title{ Hidden Quasicrystal in Hofstadter Butterfly }
\author{ Indubala I Satija }
 \affiliation{School of Physics and Astronomy and Computational Sciences, George Mason University,
 Fairfax, VA 22030 }
\date{\today}
\begin{abstract}
Topological description of hierarchical sets of spectral gaps of Hofstadter butterfly 
is found to be encoded in a quasicrystal where magnetic flux
plays the role of a phase factor that shifts the origin of the quasiperiodic order.
Revealing an intrinsic frustration at smallest energy scale, described by $\zeta=2-\sqrt{3}$, this irrational number characterizes the universal butterfly  and is related
to two quantum numbers that includes the Chern number of quantum Hall states. 
With a periodic drive that induces phase transitions in the system,
the fine structure of the butterfly is shown to be amplified
making states with large topological invariants  accessible experimentally .
\end{abstract}
\pacs{03.75.Ss,03.75.Mn,42.50.Lc,73.43.Nq}
\maketitle

Hofstadter butterfly\cite{Hof}, also known as Gplot\cite{Gplot} is a fascinating two-dimensional spectral landscape, a quantum fractal
 where energy gaps
encode topological quantum numbers associated with the Hall conductivity\cite{QHE,TKNN}. 
This intricate mix of order and complexity is due to two competing periodicities
in a crystalline lattice subjected to a magnetic field.
The allowed energies in the spectrum of such a system are discontinuous function of the magnetic flux penetrating the unit cell, while the gaps are continuous except at discrete points. 
The stunning smoothness of spectral gaps may be traced to topology which makes spectral properties stable with respect to small fluctuations in the magnetic flux.
The Gplot continues to arouse a great deal of excitement since its discovery,
and there are various recent attempts to capture this iconic spectrum in laboratory\cite{Moire}.

Fractal properties of the butterfly spectrum have been the subject of various theoretical studies\cite{fractal,Wil,fractal1,goldman}. However,
detailed description quantifying self-similar universal properties of the butterfly fractal has not been reported previously.
In this paper, we present a geometrical framework to decode the nesting
rule that will reproduce the entire butterfly landscape, as one zooms in to the Gplot, and obtain universal scaling that characterizes 
the spectrum near half-filling.
We address the question of both spectral and topological universality. The spectral gaps are labeled by two
quantum numbers which we denote as denote as $\sigma$ and $\tau$. The integer $\sigma$ is the Chern number 
, the quantum number associated with Hall conductivity\cite{TKNN} and 
$\tau$ is an integer. These quantum numbers satisfy the Diophantine equation (DE)\cite{Dana},
\begin{equation}
\rho= \phi \sigma +\tau
\label{Dio}
\end{equation}
where $\rho$ is the particle density when Fermi level is in the gap and $\phi$ denotes the magnetic flux per unit cell.

\begin{figure}[htbp]
\includegraphics[width = 1\linewidth,height=1\linewidth]{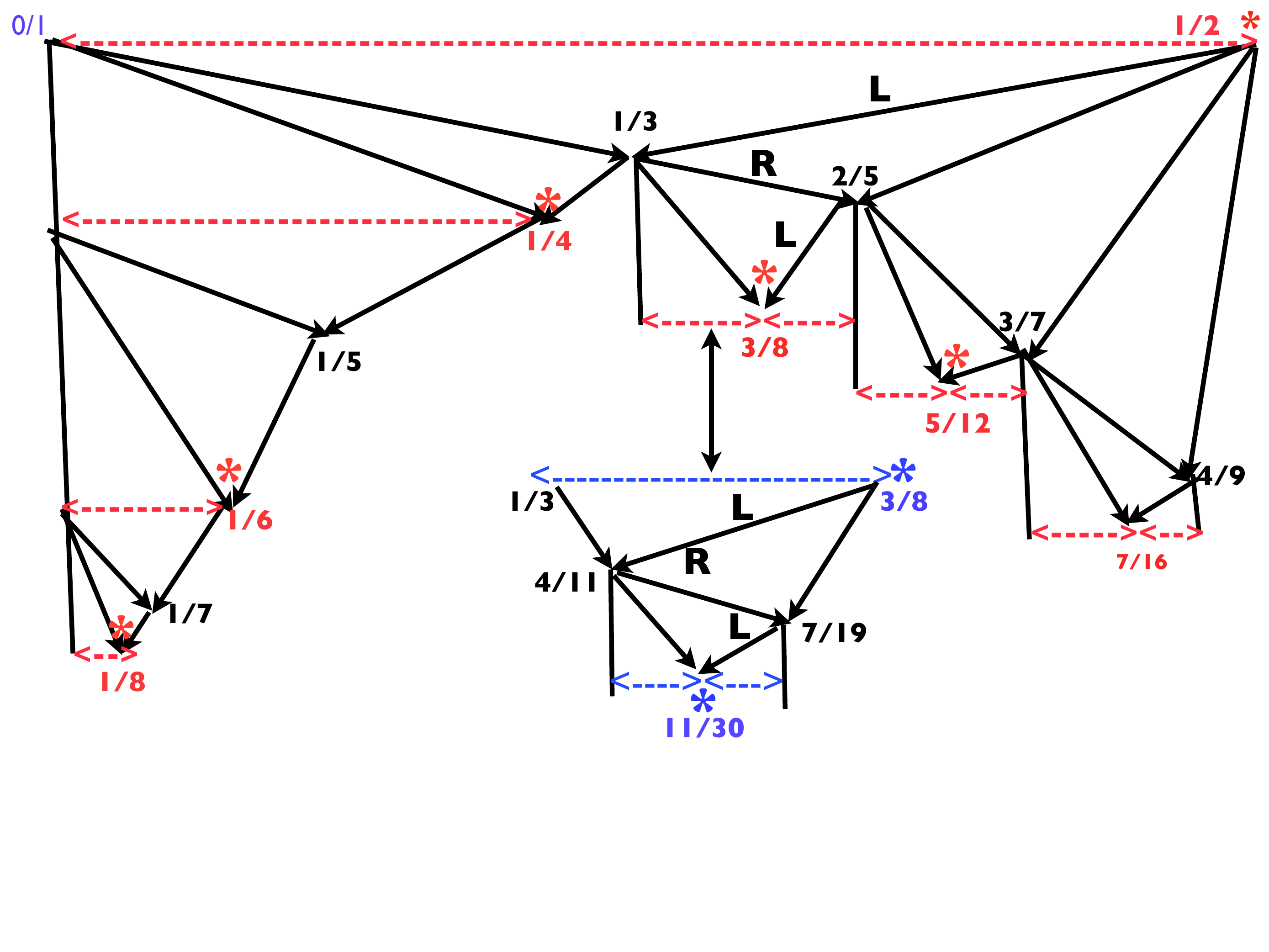}
\leavevmode \caption{ Farey tree based division of the level-$0$ interval [0,1/2]  into sub-intervals, shown with
a dashed (red) line with arrows. Each sub-interval contains the entire butterfly spectrum where * show the centers of the butterfly from where
the Farey path "LRL" begins and also ends. Points where two arrows meet represent the Farey sum\cite{Fsum} of two rationals located at the starting points of the two arrows.
The vertical double arrow
shows the levels $2$ construction starting with level-$1$ interval [$1/3 , 2/5]$.  We note that there are many other possible intervals to begin the recursive process, that are not  explicitly shown in the figure,
such as $[1/4,1/3]$ or $[1/6,1/5$].}
\label{HBFarey}
\end{figure}

Somewhat reminiscent of well known geometrical fractals, nesting property of the Gplot is found to be embedded in a
Farey tree\cite{Farey} organization of various rational magnetic flux values ( See Fig. (\ref{HBFarey})).  As we zoom in to the 
butterfly spectrum, where magnetic flux intervals
shrink by $\zeta^2$ ($\zeta=2-\sqrt{3}$), we find an algebraic recursive rule for $\sigma$ and $\tau$. These quantum numbers form a generalized Fibonacci sequence ( Eq. (\ref{Crecur})) associated with rational approximants  of $\zeta$, which has period-$2$ continued fraction expansion,
determining both $\sigma$ and $\tau$ at finer and finer scales in the butterfly fractal. This
signals a kind of {\it topological frustration} that may be responsible for magnificently complex structure of the butterfly.
This hidden quasicrystal, which we will refer 
as Hofstadter-Chern lattice, describes the entire butterfly landscape near half-filling. Here magnetic flux
simply plays the role of a phase factor that leaves quasiperiodic topological pattern unchanged.

\begin{figure}[htbp]
\includegraphics[width = 1.1\linewidth,height=1.1\linewidth]{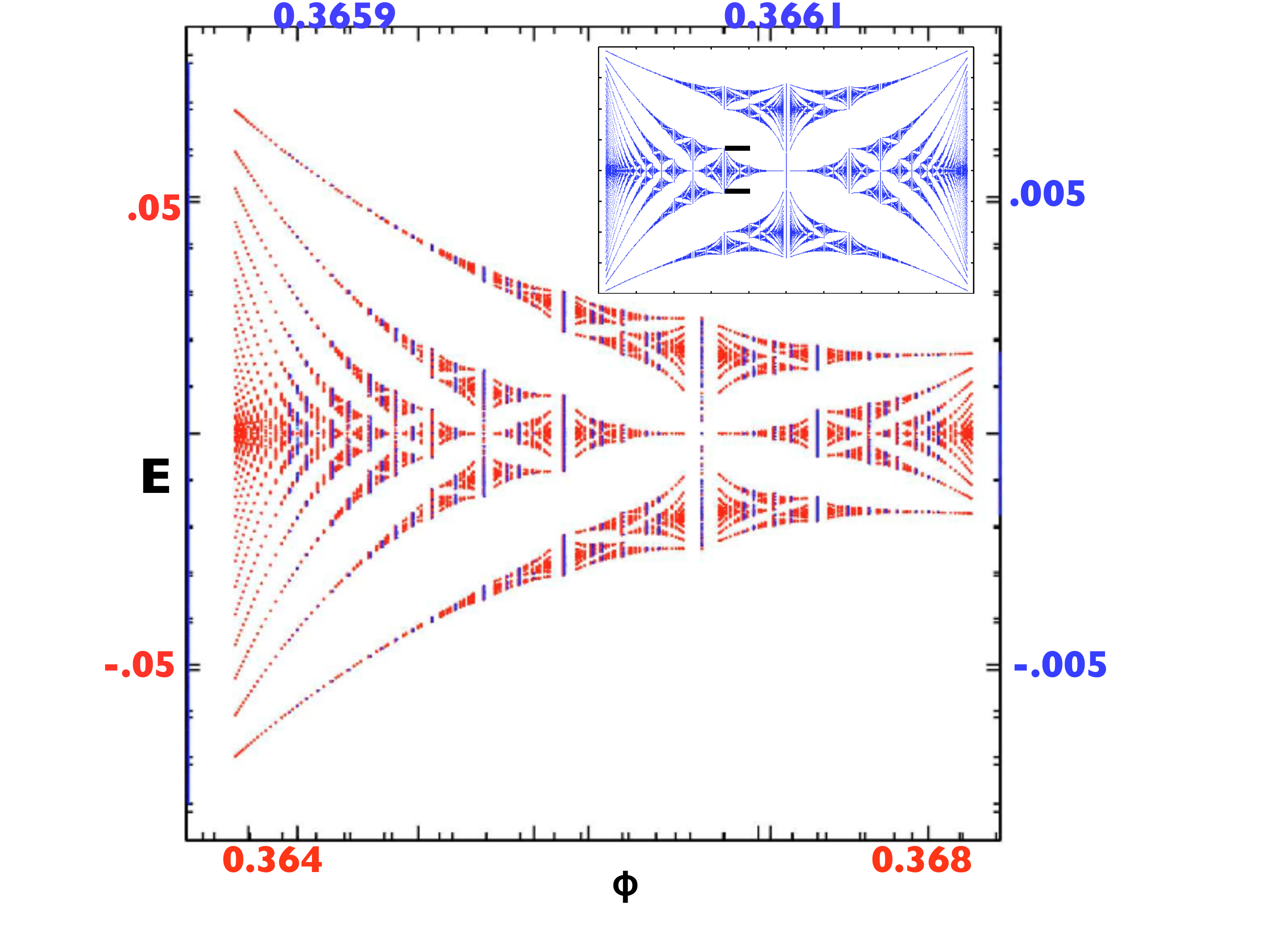}
\leavevmode \caption{(color on line) 
The inset shows level-$0$ butterfly where the two horizontal bars show the level-$1$ interval $[1/3,2/5]$. The corresponding level-$2$ and level $3$ butterflies in $\phi$ intervals $[4/11,7/19]$   and
$[15/41,26/71]$ overlay, giving energy scale ratio $\approx 10$ as shown in red and blue respectively. This illustrates an approach to fixed point butterfly fractal as $E$ and $\phi$ intervals shrink to zero.}
\label{butt}
\end{figure}

In contrast to geometrical description of
topological universality, the universal spectral property of the Gplot 
is obtained numerically. Corresponding to a scaling ratio of $\zeta^2$ for the magnetic flux interval, and $\zeta$ for quantum numbers $\sigma$ and also $\tau$,
 the spectrum is found to scale approximately as $\zeta^{\sqrt{3}}$.
We obtain the two-dimensional fixed point fractal ( See Fig. (\ref{butt})) that characterizes the entire landscape near half-filling
and verify its universality as magnetic flux varies.

In our investigation of the fractal properties of
the Hofstadter butterfly, one of the key guiding concepts is
a corollary of the DE equation that quantifies the topology of the fine structure near rational fluxes. 
We show that, for every rational flux, an
infinity of possible solutions of the DE, although not supported in the simple square lattice model
, are present in close vicinity of the flux.
Consequently, perturbations that induce topological phase transitions can transform tiny gaps with large topological
quantum numbers into major gaps. This might facilitate the creation of such states in an experimental setting.
We illustrate this amplification by periodically driving the system. 

Model system we study here consists of (spinless) fermions in a square lattice. 
Each site is labeled by a vector ${\bf r}=n\hat{x}+m\hat{y}$, where $n$, $m$ are
integers, $\hat{x}$ ($\hat{y}$) is the unit vector in the $x$ ($y$) direction, and $a$ is the lattice spacing. The tight binding Hamiltonian has the form
\begin{equation}
H=-J_x\sum_{\bf r}|\mathbf{r}+\hat{x} \rangle\langle \mathbf{r}|
-J_y\sum_{\bf r}|\mathbf{r}+\hat{y} \rangle e^{i2\pi n\phi} \langle \mathbf{r}|
+h.c. \label{qh}
\end{equation}
Here, $|\mathbf{r}\rangle$ is the Wannier state localized at site $\mathbf{r}$. $J_x$ ($J_y$)
is the nearest neighbor hopping along the $x$ ($y$) direction.
With a uniform magnetic field $B$ along the $z$ direction,
the flux per plaquette, in units of
the flux quantum $\Phi_0$, is $\phi=-Ba^2/\Phi_0$. 
Field $B$ gives
rise to the Peierls phase factor $e^{i2\pi n\phi}$ in the hopping. 

In the Landau gauge realized in experiments\cite{Ian}, the vector potential
$A_x=0$ and $A_y = -\phi x$, the Hamiltonian is cyclic in $y$ so
the eigenstates of the system
can be written as $\Psi_{n,m}= e^{ik_y m} \psi_n $ where $\psi_n$ satisfies the Harper equation\cite{Harper}
\begin{equation}
e^{ik_x}\psi^r_{n+1}+e^{-ik_x} \psi^r_{n-1} + 2\lambda \cos ( 2 \pi n \phi+ k_y)\psi^r_n = E \psi^r_n .
\label{harper}
\end{equation}
Here $n$ ($m$) is the site index along the $x$ ($y$) direction, $\lambda=J_y/J_x$
and $\psi^r_{n+q} =\psi^r_n$, $r=1, 2, ...q$ are linearly independent solutions.
In this gauge the magnetic Brillouin zone
is $ -\pi/qa \le k_x \le \pi/qa$ and $-\pi \le k_y \le \pi$.

At flux $\phi=p/q$, the energy spectrum has $q-1$ gaps.
For Fermi level inside each energy gap, the system
is in an integer quantum Hall state\cite{QHE} characterized by its Chern number
$\sigma$ which gives transverse conductivity $C_{xy}=\sigma {e^2}/{h}$\cite{TKNN}.
The Chern number $\sigma$ and an integer $\tau$ label various gaps of the butterfly and are the solutions of DE\cite{Dana},
with possible values,
\begin{equation}
(\sigma, \tau ) = (\sigma_0-n q, \tau_0 - np )
\label{DEsol}
\end{equation}
Here $\sigma_0, \tau_0$ are any two integers that satisfy the Eq. (\ref{Dio}) and $n$ is an integer.
The quantum numbers $\sigma$ that determines the quantized Hall conductivity corresponds to
the change in density of states when the magnetic flux quanta in the system 
is increased by one and whereas the quantum number $\tau$ is the change in density of states
when the period of the potential is changed so that there is one more unit cell in the system\cite{fractal1}.

For any value of the magnetic flux , the system described by the Hamiltonian (\ref{qh}), supports  only $n=0$ solution  of Eq. (\ref{DEsol}) for the
quantum numbers $\sigma$ and $\tau$. This is
due to the absence of any gap closing that is essential for topological phase transition
to states with higher values of $\sigma, \tau$. However, DE which relates continuously varying quantities $\rho$ and $\phi$
with integers  $\sigma$ and $\tau$, has some important consequences about topological changes in close vicinity of rational values of $\phi$. We now show that the infinity of solutions
depicted in Eq.(\ref{DEsol}) reside in close proximity to the flux $\phi$  and label the fine structure of the butterfly in Gplot .
In DE, we substitute $\phi = \phi_0+\delta \phi$, where $\phi_0=p_0/q_0$, and $ \rho = \rho_0 + \delta \rho$ and the
corresponding quantum numbers as $\sigma = \sigma_0+\Delta \sigma$ and $\tau = \tau_0 +\Delta \tau$. 
Now, taking the limits as $\delta \phi$ and $\delta \rho$ go to zero, we obtain,

\begin{equation}
\phi_0 \Delta \sigma+ \Delta \tau = 0;\,\,\,\,
\frac{\Delta \sigma}{\Delta \tau}= -\frac{q_0}{p_0}
\label{Dchern}
\end{equation}

Since both $\Delta \sigma$ and $\Delta \tau$ are integers and $p_0$ and $q_0$ are relatively prime, the simplest solutions of Eq. (\ref{Dchern})
are $\Delta \sigma = \pm n q_0$ and $\Delta \tau = \mp n p_0$, where $n=0,1,2,...$. These solutions describe
the fine structure of the butterfly near $\phi_0$.
Consequently, the spectral gaps near $\phi=1/q$ have
Chern numbers changing by a multiple of $q$. This
suggests a semiclassical picture near $\phi=p/q$ in terms of an effective Landau level
theory with cyclotron frequency renormalized by $q$.

Another important consequence of DE near half-filling is a relationship between the magnetic flux at the center of the butterfly
and the topological integer that labels the butterfly.
For $E=0$, rational values of magnetic flux  with even denominators result in 
a set of $4$-swaths converging to a structure resembling a
butterfly. Although, precisely at the center of the butterfly, the Chern number is undefined, the structure emanating from the center can be
associated with a unique value of $\sigma$ and $\tau$ as described below.
For energy gaps near $E=0$, we write $\rho=1/2+\epsilon$ and $\phi = p_c/q_c  + \delta$.
Substituting  these values of $\rho$ and $\phi$ in  DE  and taking the limit $\epsilon, \delta \rightarrow 0$, we obtain
topological integers that characterize the  four wings of the butterfly.

\begin{equation}
\sigma= \pm\frac{q_c}{2};\,\,\,\,
\tau=\frac{|p_c \mp 1|}{2}
\label{CD}
\end{equation}

The Chern number $\sigma$ changes sign in $4$-wings of the butterfly and $\tau$ values differ by unity in the left and the right wing. 
We label each butterfly uniquely by the magnitude of $\sigma$ and
$\beta=2|\tau|+1$ .

We now address the question of recursive scheme for topological invariants for the nested sets of butterflies that are centered at $E=0$
and seek a universal fixed point function that underlies this entire spectral landscape near half-filling.
We note  that the off-centered
butterflies appear as the continuation of the central butterflies, with discontinuities at discrete set of points.

To describe the
hierarchical structure of the butterfly fractal, we introduce a notion of "levels" where higher levels correspond to viewing the Gplot at smaller and smaller scale
in $E$ vs $\phi$ plot. At level-$0$ we have the central butterfly in the $\phi$-interval $[0,1]$ with center at $\phi_c=1/2$.
In the interval, $[0,1/2]$,
, there is a colony of asymmetrical butterflies
all meeting at the left boundary $\phi=0$, as shown in Fig (\ref{butt}).
Analogously, there is also a right colony in the $\phi$ interval $[1/2,1]$ sharing the right boundary $\phi=1$.
Without loss of generality we will focus on the spectrum only in the $\phi$-interval $[0, 1/2$] consisting 
of set of butterflies centered at $\phi_c=\frac{1}{2m}$, where $m=1,2,3...$. From Eq. (\ref{CD}),
these butterflies are labeled with quantum numbers $\sigma=m$ and $\beta=1$. 
At any level $l$, the
central butterfly, with center at flux $f_c(l)=\frac{p_c(l)}{q_c(l)}$ is confined between its
left and right boundaries at flux values $f_L(l)=\frac{p_L(l)}{q_L(l)}$ and $f_R(l)=\frac{p_R(l)}{q_R(l)}$ respectively. 

A close inspection of the Gplot reveals that Farey sequences are the key to systematically sub-divide the $\phi$ interval, where each new interval reproduces the entire butterfly.
We illustrate this process in the Fig. (\ref{HBFarey}) where a specific Farey path represented symbolically as "LRL"
is essential in constructing the recursive scheme. 
It turns out that the choice of the initial interval is irrelevant as the universal scaling properties are independent of $\phi$ .
We note that beyond level-$0$, butterflies do not exhibit reflection symmetry about their centers ( See Fig. (\ref{butt}) , however,
the recursive scheme for the left and the right intervals are identical.
Figure shows two sets of intervals:
(I) an infinite set constructed to the right of $f_L$ where each Farey sum\cite{Fsum} includes $f_L$ and (II)
an infinite set constructed to the left of $f_c$ where each Farey sum includes $f_c$.
The recursive rules for constructing
the left and right boundaries and the center of the butterfly from level $l$ to level $l+1$ are given by,
\begin{eqnarray*}
f_L(l+1)&=&f_L(l) \bigoplus f_c(l)\\
f_R(l+1)&=&f_L(l+1) \bigoplus f_c(l)\\
f_c(l+1)&=&f_L(l) \bigoplus f_R(l)
\label{RR}
\end{eqnarray*}
At a level-$l$ of the nesting scheme, the quantum numbers $\sigma, \tau$ are found to scale by rational approximants of $\zeta$, obtained by truncating its continued fraction expansion, 
\begin{equation}
 \zeta = [3,1,2,1,2,1,2....]= \cfrac{1}{3
          + \cfrac{1}{1
          +\cfrac{1}{2
          + \cfrac{1}{1 + \cfrac{1}{2.....} } } }}
          \label{cont}
\end{equation}

Denoting $l^{th}$ rational approximant of $\zeta$ as $\zeta(l) = P(l)/Q(l)$, the quantum numbers (using Eqs (\ref{CD},\ref{RR})) for the hierarchical set beginning with the interval
 $[1/3, 2/5]$ are, $\sigma(l) = Q(2l)$, $\beta(l) = Q(2l-1)$ and
 $\sigma(l-1) = P(2l)$ , $\beta(l-1) = P(2l-1)$. 
 This reveals a simple process of two independent integers described by a single irrational with period-$2$ continued fraction as
  integers related to even(odd) approximants determine $\sigma$($\beta$). 
 Using number theoretical properties of $\zeta$, it can be shown that a single recursion, where $\zeta(l)$ is evaluated at every \underline{second} iteration, describes the recursive scheme for both $\sigma$ and $\beta$,
\begin{equation}
A(l+1)=4A(l)-A(l-1)
\label {Crecur}
\end{equation}
where $A=\sigma, \beta$.  Interestingly, this recursion determines the quantum numbers $\sigma, \tau$
for all values of $\phi$, with 
the initial values in Eq. (\ref{Crecur}) determined by the choice of $\phi$ interval. Asymptotically, $\sigma(l) \rightarrow \zeta^{-l}$, $\tau \rightarrow \zeta^{-l}$ 
and the underlying $\phi$ interval
scales as, $\phi \rightarrow \zeta^{2l}$.

For the butterfly fractal shown in Fig. (\ref{butt}), the entire band spectrum is numerically found to scale approximately as,
$\Delta E \approx 10^{l} \approx \zeta^{\sqrt{3}l}$.
Although the precise value of quantum numbers ( and hence the universal butterfly fractals) depend upon $\phi$, the scaling ratios between two successive levels
is $\phi$ independent.
We summarize the universal scaling ratio $R_x$ where $x=\phi, \sigma, \tau, E$,
\begin{equation}
R_{\phi} = \zeta^{-2};\,\,\,
R_{\sigma}=R_{\tau} = \zeta;\,\,\,
R_E \approx \zeta^{\sqrt{3}}
\end{equation}
Therefore, topological variations occur at a slower rate than the corresponding spectral variations as one views the butterfly at smaller and smaller scale.
In summary, the topology of the universal butterfly is found to be encoded in a single irrational number that
due to its period-$2$ continued fraction with entries $1$ and $2$ is the simplest possible number that underlies two
distinct quantum numbers that appear in DE.  The quasiperiodic Hofstadter Chern -lattice generated by $\sigma(l), \tau(l)$ 
is a generalization of the well known Fibonacci lattice,
characterized by the nesting property that each generation is the sum of two previous generations.
 The magnetic flux simply plays the role of a phase factor, an additional degree of freedom
for the quasiperiodic Chern-lattice\cite{QP}, that shifts the origin of the quasi-periodic order.

\begin{figure}[htbp]
\includegraphics[width = 1.3\linewidth,height=1.3\linewidth]{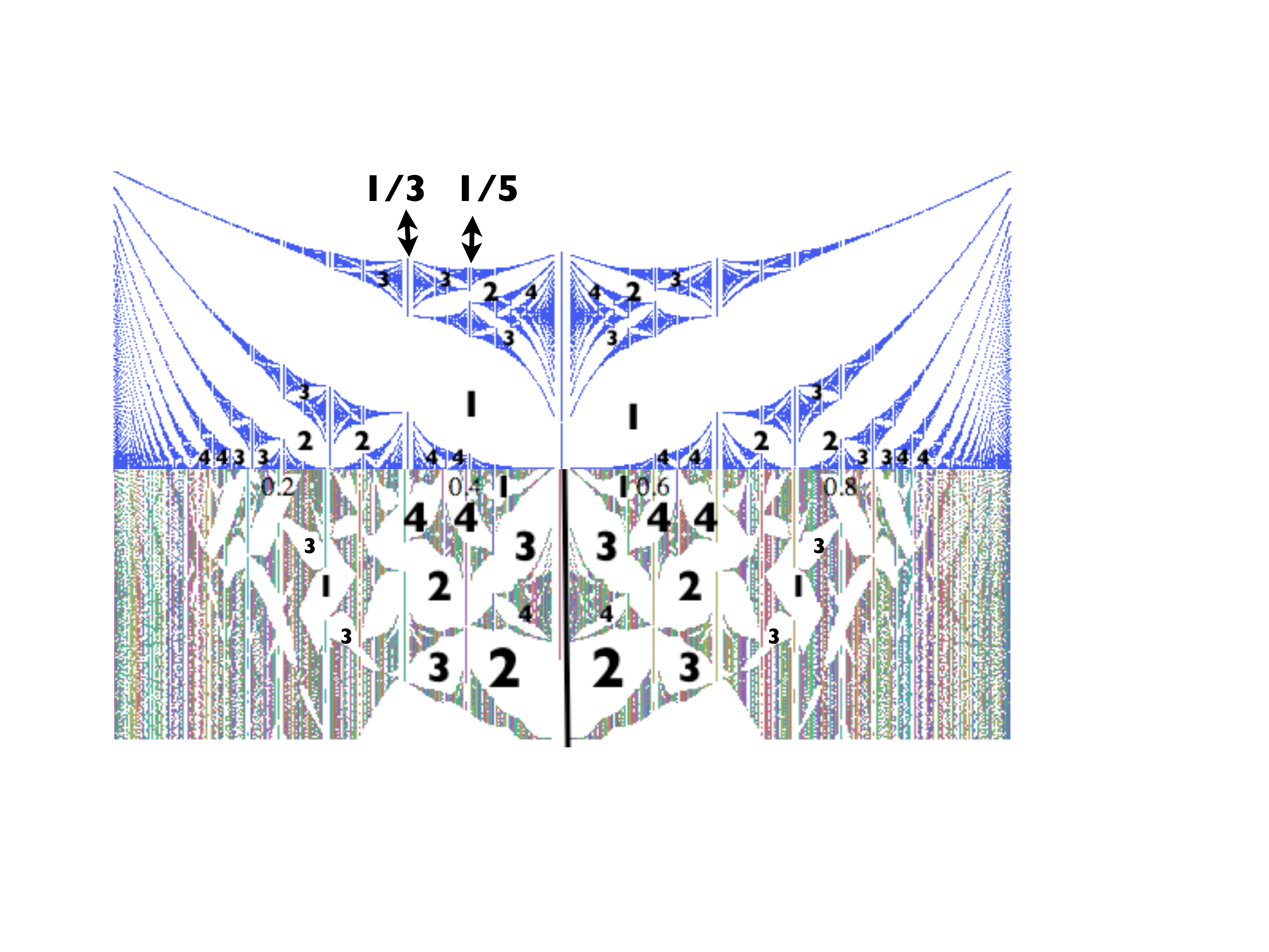}
\leavevmode \caption{(color on line) Upper half of the butterfly spectrum for static ($J_x=J_y$)(upper)
and lower-half of the quasienergy spectrum for driven system with $\bar{J}=.33,\bar{\lambda}=1.1$ (lower).
Dominance of higher Chern states in kicked system is due to
phase transitions where Chern-$1$ state for $\phi=1/3$ 
is transformed
into Chern $-2(=1-3)$ state while Chern ($-2$, $1$) states of $\phi=2/5$ 
evolve into Chern $3(=-2+5)$,$-4(=1-5)$ states.}\label{KickB}
\end{figure}

We next address the question of physical relevance of states of higher topological numbers in view of 
the fact that size of the spectral gaps decreases exponentially with $\sigma$, as confirmed by our
numerical study of the system described by Eq. (\ref{harper}).
We now show that by perturbing such systems, we can induce quantum phase transitions to topological states
with $n > 0$ given by (\ref{DEsol}) with dominant gaps characterized by higher Chern numbers. We study butterfly spectrum for a
periodically kicked quantum Hall system\cite{MLprl}
where $J_y$ is a periodic function of time $t$ with period-$T$\cite{MLprl},
\begin{equation}
J_y= \lambda \sum_n \delta(t/T-n)
\end{equation}
The time evolution operator of the system, defined by $|\psi(t)\rangle=U(t)|\psi(0)\rangle$,
has the formal solution $U(t)={\cal T}\exp[-i\int^t_0H(t')dt']$,
where ${\cal T}$ denotes time-ordering and we set $\hbar=1$ throughout.
The discrete
translation symmetry $H(t)=H(t+T)$ leads to a convenient basis
$\{|\phi_\ell\rangle\}$, defined as the eigenmodes of Floquet operator $U(T)$,
\[
U(T)|\phi_\ell\rangle=e^{-i\omega_\ell T}|\phi_\ell\rangle.
\]
We have two independent
driving parameters,
$\bar{J}=J_x T/\hbar$ and $\bar{\lambda}=\lambda T/\hbar$.
For rational flux $\phi=p/q$, 
$U$ is a $q\times q$ matrix with
$q$ quasienergy bands that reduce to the energy bands of the static system as $T \rightarrow 0$.

New topological landscape of the driven system as shown in the Fig. (\ref{KickB}) can be understood by determining the topological states
of flux values corresponding to simple rationals such as $1/3$, $2/5$.  In the Fig. (\ref{KickB}), parameter values correspond to the case
where the Chern-$1$ gap associated with $1/3$ has undergone quantum phase transition to a $n=1$ solution of the DE (Eq. (\ref{DEsol})) and
the Chern-$-2,1$ states of $2/5$ have also undergone transitions to Chern-$3,-4$ state. This almost wipes out
the Chern-$1$ state from the landscape, exposing the topological states of higher Cherns that existed in tiny gaps in the static system.
Gap amplifications for Chern $2$, $3$ and $4$ states in periodically driven quantum hall system
may provide a possible pathway to see fractal aspects of
Hofstadter butterfly and engineer states with large Chern numbers experimentally.

Recently, there is renewed interest in quasiperiodic systems\cite{QP, SN,Recent1, Recent2, Recent3} due to
their exotic characteristics that includes their relationship
to topological insulators. Hofstadter-Chern lattice is 
 intrinsically frustrated system, at smallest energy scale induced by inherent frustration created by the magnetic flux $\phi$.
This suggest a new source of recursive behavior whose
deeper understanding and significance  needs further investigation.


\begin{thebibliography}{99}

\bibitem{Hof} D. Hofstadter, Phys Rev B, {\bf 14} (1976) 2239.

\bibitem{Gplot} D. Hofstadter, " Godel, Escher and Bach", Vintage Books Edition 1989. 

\bibitem{QHE} von Klitzing K, Dorda G and Pepper M , Phys. Rev. Lett {\bf 45} 494 (1980).

\bibitem{TKNN} D. J. Thouless, M. Kohmoto, M. P.
Nightingale, and M. den Nijs, Phys. Rev. Lett. {\bf 49}, 405 (1982).

\bibitem{Moire} C. R Dean et al, Nature 12186, 2013; M. Aidelsburger, Phys Rev Lett, {\bf 111} 185301 (2013);
Hirokazu Miyake et al, Phys Rev Lett, {\bf 111} 185302 (2013).

\bibitem{fractal} Wannier, G. H. , Phys. Status Solidi B 88, 757–765 (1978); MacDonald, A. Phys. Rev. B 28, 6713–6717 (1983).

\bibitem{Wil} M. Wilkinson, J. Phys. A: Math. Gen. 20 (1987)4337-4354;
J. Phys. A: Math, Gen.21 (1994) 8123-8148.

\bibitem{fractal1}
Ming-Che Chang and Qian Niu, PRB, {\bf 53} 7010, 1996.

\bibitem{goldman} N.Goldman, J. Phys B, 42, 055302, 2009.

\bibitem{Dana} Itzhack Dana, J . Phys. C: Solid State Phys. 18 (1985) L679-L683

\bibitem{Farey}Hardy and Wright (1979), " An introduction to the Theory of Numbers", Oxford University Press,Chapter III, 
\bibitem{Ian}
Y.-J. Lin \emph{et al.}, Phys. Rev. Lett. \textbf{102}, 130401 (2009);
I. B. Spielman, Phys. Rev. A \textbf{79}, 063613 (2009).
\bibitem{Harper} P.G.Harper,Proc.Phys.Soc.London Sect. A68,874(1955).

\bibitem{Fsum} The Farey sum, denoted by $\bigoplus$ between two rationals $\frac{p1}{q1}$ and $\frac{p2}{q2}$ is defined 
as $\frac{p1}{q1} \bigoplus \frac{p2}{q2}=\frac{p1+p2}{q1+q2}$.

\bibitem{MLprl} M. Lababidi, I Satija and E. Zhao, Phys Rev Lett, {\bf 112} (2014) 026805.

\bibitem{QP} Y. E. Kraus and O. Zilberberg, Phys. Rev. Lett. 109, 116404
(2012)
\bibitem{SN}  I. Satija and G. Naumis, Phys Rev B {\bf 88} 054204 (2013).
\bibitem{Recent1} S. Janecek,M. Aichinger, and E. R. Herna ́ndez, Phys Rev B {\bf 87}, 235429 (2013);
\bibitem{Recent2}I. Dana, arXiv:1310.7970, 2013
\bibitem{Recent3} Mor Verbin,Oded Zilberberg,Yoav Lahini,Yaacov E. Kraus,
and Yaron Silberberg, arXiv:1403.7124v1

\end{thebibliography}
\end{document}